\documentclass[prb,superscriptaddress,amsfonts,twocolumn,showpacs,amssymb,floatfix, bibnotes]{revtex4}
\usepackage{color}
\usepackage{graphicx}% Include figure files
\usepackage{dcolumn}% Align table columns on decimal point
\usepackage{bm}% bold math
\usepackage{mathrsfs}
\begin{document}

%\preprint{APS/123-QED}

\title{Systematic $^{75}$As NMR study of the dependence of low-lying excitations on F-doping\\ 
in the iron oxypnictide LaFeAs(O$_{1-x}$F$_{x}$) }

\author{Yusuke Nakai}
\email{nakai@scphys.kyoto-u.ac.jp}
\affiliation{Department of Physics, Graduate School of Science, Kyoto University, Kyoto 606-8502, Japan,}
\affiliation{TRIP, JST, Sanban-cho bldg., 5, Sanban-cho, Chiyoda, Tokyo 102-0075,}
\author{Shunsaku Kitagawa}
\affiliation{Department of Physics, Graduate School of Science, Kyoto University, Kyoto 606-8502, Japan,}
\affiliation{TRIP, JST, Sanban-cho bldg., 5, Sanban-cho, Chiyoda, Tokyo 102-0075,}
\author{Kenji Ishida}
\affiliation{Department of Physics, Graduate School of Science, Kyoto University, Kyoto 606-8502, Japan,}
\affiliation{TRIP, JST, Sanban-cho bldg., 5, Sanban-cho, Chiyoda, Tokyo 102-0075,}

\author{Yoichi Kamihara}
\affiliation{TRIP, JST, Sanban-cho bldg., 5, Sanban-cho, Chiyoda, Tokyo 102-0075,}

\author{Masahiro Hirano}
\affiliation{ERATO-SORST, Tokyo Institute of Technology, Yokohama 226-8503, Japan,}
\affiliation{Frontier Research Center, Tokyo Institute of Technology, Yokohama 226-8503, Japan,}

\author{Hideo Hosono}
\affiliation{ERATO-SORST, Tokyo Institute of Technology, Yokohama 226-8503, Japan,}
\affiliation{Frontier Research Center, Tokyo Institute of Technology, Yokohama 226-8503, Japan,}
\affiliation{Materials and Structures Laboratory, Tokyo Institute of Technology, Yokohama 226-8503, Japan}

%\date{October 20, 2008}

\begin{abstract}
We report systematic $^{75}$As NMR studies on LaFeAs(O$_{1-x}$F$_{x}$) ($0\le x\le0.14$). 
At $x=0.04$ near the phase boundary, from resistivity, spin-lattice relaxation rate $1/T_1$, and NMR spectrum measurements, we found weak magnetic order at $T_N\simeq 30$ K. 
Antiferromagnetic (AFM) fluctuations proved through $1/T_1$ are suppressed significantly with F-doping, and pseudogap behavior without pronounced AFM fluctuations is observed at $x=0.11$ where $T_c$ is maximum. 
This significant suppression of $1/T_1T$ upon F-doping while $T_c$ remains nearly unchanged suggests that low-energy AFM fluctuations probed with $^{75}$As NMR do not play a crucial role in the superconductivity.

\end{abstract}

%\keywords{Suggested keywords}%Use showkeys class option if keyword display desired
\maketitle

The recent discovery of the iron oxypnictide superconductor LaFeAs(O$_{1-x}$F$_x$) with $T_c=26$ K~\cite{KamiharaFeAs} has stimulated intense research on the origin of its high $T_c$. 
Superconductivity in the iron oxypnictides appears upon F-doping in close proximity to parent phases which exhibit stripe antiferromagnetic (AFM) order, along with structural transition from tetragonal to orthorhombic phases.~\cite{Cruz,Nomura,NakaiJPSJ2008} 
Thus, it is natural to investigate an interplay between superconductivity and spin fluctuations associated with the magnetic ordering. 
In addition, because electron-phonon coupling is too weak to account for the high $T_c$,~\cite{BoeriPRL2008} 
spin fluctuations due to nesting between the disconnected Fermi surfaces have been suggested to be the source of the pairing interaction.~\cite{SinghPRL2008,MazinPRL2008,KurokiPRL2008} 
F-doping, corresponding to electron doping, suppresses the nesting and thus low-lying excitations originating from the nesting-related magnetic fluctuations, however $T_c$ is relatively insensitive to F-doping.~\cite{KamiharaFeAs} 
Hence, investigations on the F-doping dependence of spin dynamics in the normal and superconducting (SC) states in LaFeAs(O$_{1-x}$F$_{x}$) will provide crucial information concerning the relationship between superconductivity and spin fluctuations. 
In our previous paper, we reported NMR studies on LaFeAs(O$_{1-x}$F$_{x}$) for $x=$ 0, 0.04 and 0.11, but it remained insufficient for revealing systematic variation in spin dynamics.~\cite{NakaiJPSJ2008} 
Here, we report systematic studies on LaFeAs(O$_{1-x}$F$_{x}$) through $^{75}$As NMR in order to fully elucidate the nature of spin dynamics in a wider F-doping range. 

Polycrystalline samples of LaFeAs(O$_{1-x}$F$_{x}$) ($x=$ 0, 0.04, 0.07, 0.11, 0.14) synthesized through solid-state reaction~\cite{KamiharaFeAs} were ground into powder for NMR measurements; powder X-ray diffraction measurements indicate that the samples are mostly single-phase.~\cite{NoteCharacterization} 
The value of $x$ was estimated from the lattice constants using the Vegard's volume rule.~\cite{Vegard,KimKamiharaJPSJ} 
Electrical resistivity measurements were performed with a four-probe technique. 
From the zero-resistivity temperature in $H=0$ (see Fig.~1~(a)), the $T_c$'s were determined to be 16.3 K, 22.5 K, 22.5 K, and 12.5 K for $x=0.04$, 0.07, 0.11, and 0.14, respectively. 
A standard spin-echo technique was used for obtaining NMR spectra. 
The $^{75}$As nuclear spin-lattice relaxation rate $1/T_1$ was obtained by fitting the time dependence of the nuclear magnetization recovery after a saturation pulse. $1/T_1$ was measured at the lower peak [corresponding to $H\parallel ab$ denoted by the arrow in Fig.~1(b)] of the central transition in $H\simeq9.89$ T at 72.1 MHz for $0.04\le x\le0.14$, and, for $x=0$, in $H\simeq5.49$ T at 40.5 MHz. 
%************************Fig.1***************************************
\begin{figure}
\begin{center}
\includegraphics[width=7.5cm]{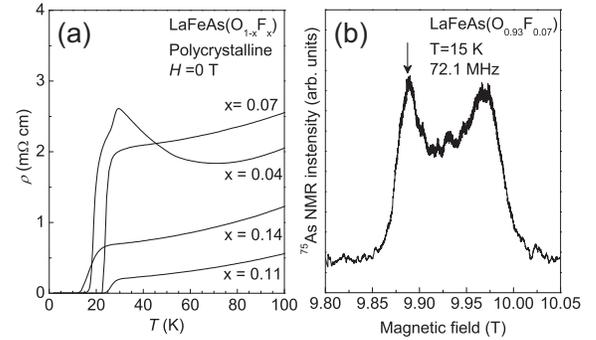}
\end{center}
\caption{(a)$T$-dependence of resistivity in LaFeAs(O$_{1-x}$F$_x$). (b)The center peak of $^{75}$As NMR spectrum at $x=0.07$. The arrow denotes the peak corresponding to $H\parallel ab$.}
\end{figure}
%***********************************************************************

The $x=0.04$ sample (LaFeAs(O$_{0.96}$F$_{0.04}$)) is located near the boundary between the AFM and SC phases. 
Figure~2 displays the $T$-dependence of the resistivity $\rho$, $1/T_1$ of $^{75}$As, and the full width at half maximum (FWHM) of the $^{75}$As NMR spectra at $x=0.04$. 
Although Fig.~1(a) indicates that the resistivity of the $x\ge$ 0.07 samples exhibits metallic behavior ($\rho\propto T^2$ below 150 K), 
the resistivity at $x=0.04$ starts to increase below 70 K, and drops below 30 K (Fig.~2(a)).
$T_1$ exhibits a single component above 30 K, and the $T$-dependence of $(T_1T)^{-1}$ follows the Curie-Weiss law $(T_1T)^{-1}\propto\frac{1}{T+\theta}$ with $\theta=10.3\pm2$ K between 30 K and 200 K,~\cite{NakaiJPSJ2008} which is characteristic of the development of AFM fluctuations. A short $T_1$ component appears below 30 K, and the longer $T_1$ component, which exhibits a superconducting anomaly at $T_c$, is plotted in Fig.~2(b). 
$1/T_1$ decreases abruptly below 30 K and then superconductivity occurs below $T_c\simeq16$ K. 
The anomaly at 30 K cannot be ascribed to the occurrence of superconductivity because ac susceptibility does not exhibit any anomaly near 30 K (not shown). 
Because the line width increases gradually below $T_N\simeq$30 K (Fig.~2(c,d)), this is attributed to weak static magnetic ordering. 
Much smaller broadening of the $^{75}$As NMR spectra than that of undoped LaFeAsO~\cite{NakaiJPSJ2008,MukudaJPSJ08} indicates a very small ordered moment for $x=0.04$. 
In LaFeAsO, we found a divergence of $1/T_1$ at $T_N\simeq142$ K due to the occurrence of stripe AFM ordering. 
In contrast, at $x=0.04$, we did not find a peak of $1/T_1$ but just a decrease of $1/T_1$, indicating that the magnetic anomaly weakens upon F-doping. 
As noted above, there emerges a short component of $1/T_1$ below 30 K whose fraction increases gradually with decreasing temperature up to $\sim$65 \% at 3 K. 
Since it is difficult to attribute such a large contribution to inhomogeneous F concentration,
this distribution of $T_1$ implies phase separation into the magnetically ordered and SC phases.~\cite{Note} 
Indeed, recent $\mu$SR measurements suggest the presence of phase separation into SC and spin-glass like phases in LaFeAs(O$_{0.94}$F$_{0.06}$) near the phase boundary.~\cite{TakeshitaJPSJ08} 
Our result is consistent with this $\mu$SR experiment; however, a definitive conclusion regarding microscopic coexistence and magnetic structure cannot be drawn due to the use of a polycrystalline sample. 
%************************Fig.2***************************************
\begin{figure}
\begin{center}
\includegraphics[width=6.5cm]{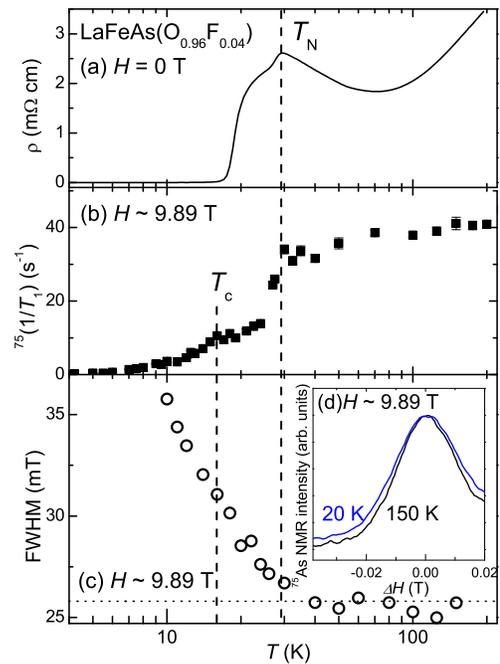}
\end{center}
\caption{(Color online) $T$-dependence of (a) resistivity, (b) $^{75}$As $1/T_1$, (c) the FWHM of the $^{75}$As NMR spectra 
and (d) $^{75}$As NMR spectrum for $H\parallel ab$ in LaFeAs(O$_{0.96}$F$_{0.04}$). The spectrum at 20 K is shifted to overlap the spectral peak at 150 K.}
\end{figure}
%***********************************************************************
%************************Fig.3***************************************
\begin{figure}
\begin{center}
\includegraphics[width=6.5cm]{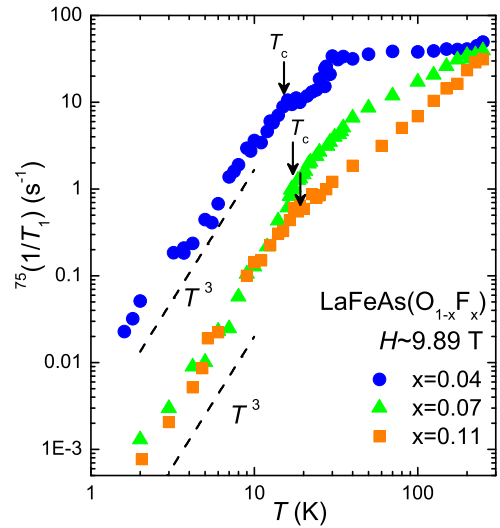}
\end{center}
\caption{(Color online) $T$-dependence of $1/T_1$ measured at the peak corresponding to $H\parallel ab$.} 
\end{figure}
%***********************************************************************

The F-doping evolution of the SC features in LaFeAs(O$_{1-x}$F$_{x}$) is shown in Fig.~3, where we present the $T$-dependence of $^{75}$As $1/T_1$ for $x=0.04$, 0.07, and 0.11, measured in $H\simeq$ 9.89 T in the $ab$-plane (the $x=0.14$ sample does not exhibit superconductivity in $H\simeq9.89$ T as was expected from a broad SC transition observed through specific-heat measurements on a sample from the same batch~\cite{KohamaFeAsPRB2008}). 
We found that $1/T_1$ for $x=0.04$, 0.07, 0.11 decreases just below $T_c$ without showing a Hebel-Slichter coherence peak and follows a $T^3$ dependence in the SC state. 
This $T^3$ dependence is observed even in $x=0.04$, in which superconductivity sets in at $T_c\simeq16$ K below $T_N\simeq30$ K. 
The robust $T^3$ dependence of $1/T_1$ at $x=0.04$ provides further evidence of phase separation into the magnetically ordered and superconducting phases, since magnetic fluctuations associated with the magnetic ordering would change the $T^3$ behavior if the magnetic order and superconductivity coexist microscopically.
Although recent ARPES~\cite{HDingBaKFe2As2} and penetration depth measurements~\cite{HashimotoPrFeAsO1-y} indicate nodeless SC-gap(s), the nodeless gap state is incompatible with the $1/T_1\propto T^3$ behavior and the lack of a Hebel-Slichter coherence peak. 
Alternatively, recent theoretical studies indicate that the lack of a coherence peak and the $T^3$ dependence of $1/T_1$ can be understood in terms of a fully gapped $s_{\pm}$ state with impurity effects.~\cite{ParkerPRB2008,ChubukovPRB2008,ParishPRB2008,Bang,NagaiNJP2008} 
We note that, however, further measurements on high-quality samples will be crucial to verify these scenarios, since $1/T_1$ should decrease exponentially well below $T_c$ in the clean limit.

In the $s_{\pm}$-wave picture, nesting-related magnetic fluctuations with the wave vector $\bf{q}_{\rm stripe}=$($\pi$, 0), (0, $\pi$) originating from the disconnected Fermi surfaces are important for superconductivity in LaFeAs(O$_{1-x}$F$_x$). 
When investigating $\bf{q}$-dependent spin dynamics using $^{75}$As NMR, one must be cautious of geometrical cancellation of magnetic fluctuations at the As site, because the As (also La) sites are located above and below the center of the Fe square lattice. 
Kitagawa {\it et al.} reported a model for the hyperfine field at the As site in terms of anisotropic hyperfine couplings between the local Fe moments and the As nucleus, and showed that $^{75}$As NMR can detect the stripe AFM order.~\cite{KitagawaBaFe2As2} 
This result is consistent with the observation via $^{75}$As and $^{139}$La NMR in LaFeAsO~\cite{NakaiJPSJ2008,MukudaJPSJ08} and BaFe$_2$As$_2$~\cite{KitagawaBaFe2As2,FukazawaJPSJ08,BaekBaFe2As2} of a divergence in $1/T_1$ originating from the stripe AFM order.

A systematic doping evolution of spin dynamics in the normal state is observed in the $T$-dependence of $^{75}$As $(T_1T)^{-1}$ in LaFeAs(O$_{1-x}$F$_x$) as shown in Fig.~4. 
In undoped LaFeAsO, a clear critical slowing down due to the AFM ordering with $\bf{q}_{\rm stripe}$ is observed at 142 K. 
For $x=0.04$, $(T_1T)^{-1}$, which is the sum of low-lying dynamical susceptibility $\chi({\bf q})$ over the Brillouin zone, follows a Curie-Weiss temperature dependence down to 30 K. In contrast, the magnetic susceptibility $\chi({\bf q}=0)$ of LaFeAs(O$_{1-x}$F$_x$) decreases with decreasing temperature.~\cite{KlingelerLaFeAsOF} Their contrasting behavior is a clear indication of the development of AFM fluctuations away from ${\bf q}=0$ at $x=0.04$. 
At $x=0.07$, $(T_1T)^{-1}$ remains nearly constant down to $T^{*}\simeq$ 40 K, then decreases rapidly. 
$T^*$ can be ascribed neither to a magnetic anomaly nor to a SC transition; invariant $^{75}$As NMR spectra rules out the former, and the absence of Meissner signal excludes the latter (not shown). 
The reduction of $(T_1T)^{-1}$ below $T^{*}$, approximately 20 K higher than $T_c$, is reminiscent of the pseudogap behavior observed in the cuprates.~\cite{TakigawaPRB1991} 
The pseudogap-like behavior is more pronounced for $x=0.11$ and 0.14, where $(T_1T)^{-1}$ decreases on cooling, approaching a nearly constant value. 
By fitting the data to $(T_1T)^{-1} = a + b\exp{(-\Delta_{\rm PG}/T)}$, we obtained $a=0.04$ (s$^{-1}$K$^{-1}$), $b=0.19$ (s$^{-1}$K$^{-1}$) and $\Delta_{\rm PG}=172\pm12$ K for $x=0.11$, and $a=0.012$ (s$^{-1}$K$^{-1}$), $b=0.18$ (s$^{-1}$K$^{-1}$) and $\Delta_{\rm PG}=165\pm15$ K for $x=$ 0.14, 
yielding almost the same pseudogap energies $\Delta_{\rm PG}$ for $x=0.11$ and 0.14 within experimental uncertainty. 
On the basis of our present results along with those reported previously, we generate the phase diagram for LaFeAs(O$_{1-x}$F$_x$) shown in Fig.~5. 
%************************Fig.4***************************************
\begin{figure}
\begin{center}
\includegraphics[width=7cm]{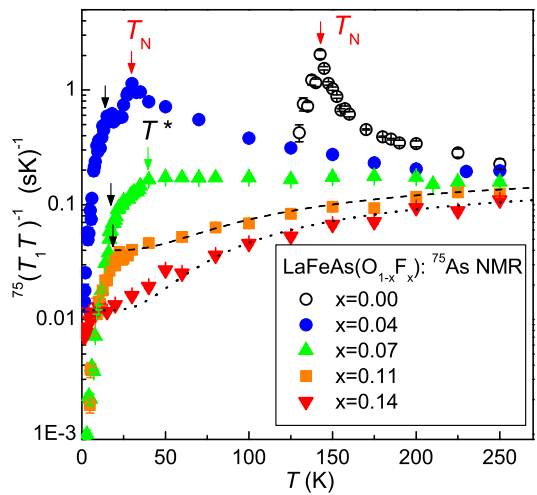}
\end{center}
\caption{(Color online) Doping dependence of $^{75}$As $(T_1T)^{-1}$. The black (green) arrows denote $T_c$ ($T^{*}$) (see text). The broken (dotted) line is a fitting to the data for $x=0.11$ (0.14). The red arrows indicate $T_N$.} 
\end{figure}
%*********************************************************************** 

The doping dependence of $(T_1T)^{-1}$ indicates that the nature of the pseudogap in LaFeAs(O$_{1-x}$F$_x$) and the cuprates differs significantly: 
(1) In the cuprates, $(T_1T)^{-1}$ decreases from temperatures well above $T_c$ and no clear anomaly is observed at $T_c$. In contrast, a clear anomaly in $(T_1T)^{-1}$ is found at $T_c$ for $x=0.11$, and Korringa behavior ($T_1T$ = const.), suggestive of a Fermi liquid state, is observed at low temperatures, which may be related to the $T^2$ behavior of the resistivity. 
Considering the multi-band nature of LaFeAs(O$_{1-x}$F$_x$), these results suggest that some Fermi surface sheets exhibit pseudogap behavior while others contribute to the Fermi liquid state. 
(2) The pseudogap behavior in LaFeAs(O$_{1-x}$F$_x$) becomes more pronounced with F-doping, opposite to the behavior observed in the cuprates. 
There, the pseudogap behavior is most pronounced near the AFM phase boundary, 
and it is possible that AFM correlations may be responsible for the pseudogap behavior. In LaFeAs(O$_{1-x}$F$_x$), however, low-energy AFM correlations are unlikely to yield the pseudogap behavior since no apparent AFM fluctuations are observed via $^{75}$As NMR for $x=$ 0.11 and 0.14. 
Furthermore, almost the same $\Delta_{\rm PG}$ is observed through $^{57}$Fe NMR measurements, suggesting that the pseudogap does not have ${\bf q}$-dependence.~\cite{TerasakiFeNMR} 
Quite recently, Ikeda suggested that the pseudogap behavior in $(T_1T)^{-1}$ may originate from band structure effects near Fermi energy.~\cite{IkedaJPSJ2008} The existence of a high density of states (DOS) just below the Fermi level, which is assigned to a hole Fermi surface around $\Gamma'$ consisting of $d_{x^2-y^2}$ orbitals in the {\it unfolded} Brillouin zone, gives rise to a $T$-dependent DOS, and the calculated $T$-dependence of $(T_1T)^{-1}$ is consistent with our results. Thus, the pseudogap behavior likely originates from its characteristic band structure in LaFeAs(O$_{1-x}$F$_x$).

Finally, we discuss the relationship between superconductivity and spin fluctuations in LaFeAs(O$_{1-x}$F$_x$). 
As shown in Fig.~4, the significant AFM fluctuations observed through $^{75}$As NMR are suppressed systematically with F-doping, and pseudogap behavior without pronounced AFM fluctuations is observed for $x=0.11$ where $T_c$ is maximum. 
We speculate that these results are attributable to the disconnected Fermi surfaces in LaFeAs(O$_{1-x}$F$_x$); 
the stripe-like AFM fluctuations originate from nesting between the hole Fermi surfaces at $\Gamma$ and electron Fermi surfaces at $M$, and the hole (electron) Fermi surfaces become smaller (larger) upon F-doping, resulting in the suppression of nesting with the wave vector $\bf{q}_{\rm stripe}$. Moreover, the pseudogap behavior appears naturally in the heavily F-doped region because the hole Fermi surface around $\Gamma'$ sinks below $E_F$ upon electron doping, producing a $T$-dependent DOS. 
Together with the recent $^{57}$Fe NMR measurements on LaFeAsO$_{0.7}$,~\cite{TerasakiFeNMR} these results suggest that low-energy magnetic fluctuations are suppressed throughout ${\bf q}$ space near optimal doping. 
This is in contrast to the cuprate superconductors, whose maximum $T_c$ occurs where the intense AFM fluctuations with $\bf{Q}=$($\pi$, $\pi$) are observed in La$_{2-x}$Sr$_x$CuO$_4$;~\cite{OhsugiJPSJ94} their AFM fluctuations are well correlated with $T_c$ and thus may be connected with the superconductivity. 
Although many theories which suggest $s_{\pm}$-wave superconductivity show the importance of nesting-related magnetic fluctuations for the SC pairing interaction in LaFeAs(O$_{1-x}$F$_x$), the reduction of $(T_1T)^{-1}$ by almost two orders of magnitude with F-doping while $T_c$ is largely unchanged suggests that the low-energy AFM fluctuations with $\bf{q}_{\rm stripe}$ observed through $^{75}$As NMR are uncorrelated with the superconductivity. 
However, it should be noted that $1/T_1$ measurements only detect low-energy magnetic fluctuations (typically $\sim$ mK order), thus we cannot exclude the possibility that magnetic fluctuations with $\bf{q}_{\rm stripe}$ persist to higher doping levels (e.g. $x=0.11$) if their characteristic energy exceeds the NMR energy window. Inelastic neutron experiments over a wide energy range would be required to fully establish the relationship between superconductivity and magnetic fluctuations. We also note that it is important to investigate the F-doping dependence of DOS for clarifying the paring interaction for superconductivity in iron pnictides, and is now in progress.
%************************Fig.5***************************************
\begin{figure}
\begin{center}
\includegraphics[width=6.5cm]{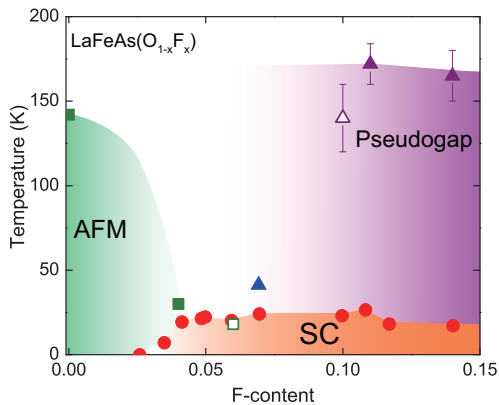}
\end{center}
\caption{(Color online) Phase diagram of LaFeAs(O$_{1-x}$F$_x$). 
The closed (open) triangle designates the pseudogap energy determined from $1/T_1$ of $^{75}$As (Knight shift of $^{19}$F cited from Ref.~31). The blue triangle indicates $T^*$ below which the pseudogap behavior was observed. 
The closed square indicates the magnetic ordering temperature $T_N$ determined from $^{139}$La and $^{75}$As NMR. 
The open square indicates the onset temperature of spin-glass like magnetism determined from $\mu$SR.~\cite{TakeshitaJPSJ08} 
$T_c$ (closed circle) is determined from the temperature where $\rho$ becomes half as that at the onset temperature.~\cite{KamiharaFeAs}}
\end{figure}
%***********************************************************************

In summary, we report systematic $^{75}$As NMR studies on LaFeAs(O$_{1-x}$F$_{x}$). At $x=0.04$ near the phase boundary, a weak magnetic anomaly occurs at $T_N\simeq 30$ K and superconductivity sets in at $T_c\simeq16$ K. 
Upon F-doping, significant AFM fluctuations observed for $x=0$ and 0.04 are suppressed systematically, 
and pseudogap behavior appears for $x=0.11$ and 0.14 without pronounced AFM fluctuations. 
The doping dependence of $(T_1T)^{-1}$ suggests that the nature of pseudogap in LaFeAs(O$_{1-x}$F$_{x}$) and the cuprates is different. 
Because $(T_1T)^{-1}$ varies drastically whereas $T_c$ is rather insensitive to F-doping, the low-energy AFM fluctuations observed through $^{75}$As NMR may be irrelevant to superconductivity. 
Understanding the drastic suppression of $(T_1T)^{-1}$ upon F-doping will be key to clarifying the source of the pairing interaction. 

We thank K.~Kitagawa, K.~Yoshimura and Y.~Maeno for experimental support and valuable discussions. 
We also thank Y.~Matsuda, T.~Shibauchi, H.~Ikeda, S.~Fujimoto and K.~Yamada for fruitful discussions. 
This work was supported by a Grant-in-Aid for the Global COE Program ``The Next Generation of Physics, Spun from Universality and Emergence" from MEXT of Japan, and by a Grants-in-Aid for Scientific Research from the Japan Society for the Promotion of Science (JSPS). Y.~N. is financially supported by JSPS.


\begin{thebibliography}{0}
\expandafter\ifx\csname natexlab\endcsname\relax\def\natexlab#1{#1}\fi
\expandafter\ifx\csname bibnamefont\endcsname\relax
  \def\bibnamefont#1{#1}\fi
\expandafter\ifx\csname bibfnamefont\endcsname\relax
  \def\bibfnamefont#1{#1}\fi
\expandafter\ifx\csname citenamefont\endcsname\relax
  \def\citenamefont#1{#1}\fi
\expandafter\ifx\csname url\endcsname\relax
  \def\url#1{\texttt{#1}}\fi
\expandafter\ifx\csname urlprefix\endcsname\relax\def\urlprefix{URL }\fi
\providecommand{\bibinfo}[2]{#2}
\providecommand{\eprint}[2][]{\url{#2}}

\end{thebibliography}


\begin{thebibliography}{31}
\expandafter\ifx\csname natexlab\endcsname\relax\def\natexlab#1{#1}\fi
\expandafter\ifx\csname bibnamefont\endcsname\relax
  \def\bibnamefont#1{#1}\fi
\expandafter\ifx\csname bibfnamefont\endcsname\relax
  \def\bibfnamefont#1{#1}\fi
\expandafter\ifx\csname citenamefont\endcsname\relax
  \def\citenamefont#1{#1}\fi
\expandafter\ifx\csname url\endcsname\relax
  \def\url#1{\texttt{#1}}\fi
\expandafter\ifx\csname urlprefix\endcsname\relax\def\urlprefix{URL }\fi
\providecommand{\bibinfo}[2]{#2}
\providecommand{\eprint}[2][]{\url{#2}}

\bibitem[{\citenamefont{Kamihara {\textit{et~al.}}}(2008)
\citenamefont{Kamihara, Watanabe, Hirano, and   Hosono}}]{KamiharaFeAs}
\bibinfo{author}{\bibfnamefont{Y.}~\bibnamefont{Kamihara\textit{ et~al.}}}, 
\bibinfo{journal}{J. Am. Chem. Soc.}
  \textbf{\bibinfo{volume}{130}}, \bibinfo{pages}{3296} (\bibinfo{year}{2008}).

\bibitem[{\citenamefont{de~la Cruz {\textit{et~al.}}}(2008)\citenamefont{de~la
  Cruz, Huang, Lynn, Li, II, Zarestky, Mook, Chen, Luo, Wang
  {\textit{et~al.}}}}]{Cruz}
\bibinfo{author}{\bibfnamefont{C.}~\bibnamefont{de~la Cruz\textit{ et~al.}}}, 
\bibinfo{journal}{Nature}
  \textbf{\bibinfo{volume}{453}}, \bibinfo{pages}{899} (\bibinfo{year}{2008}).

\bibitem[{\citenamefont{Nomura {\textit{et~al.}}}(2008)\citenamefont{Nomura,
  Kim, Kamihara, Hirano, Sushko, Kato, Takata, Shluger, and Hosono}}]{Nomura}
\bibinfo{author}{\bibfnamefont{T.}~\bibnamefont{Nomura\textit{ et~al.}}}, 
\bibinfo{journal}{Supercond. Sci. Technol.}
  \textbf{\bibinfo{volume}{21}}, \bibinfo{pages}{125028}
  (\bibinfo{year}{2008}).

\bibitem[{\citenamefont{Nakai {\textit{et~al.}}}(2008)\citenamefont{Nakai,
  Ishida, Kamihara, Hirano, and Hosono}}]{NakaiJPSJ2008}
\bibinfo{author}{\bibfnamefont{Y.}~\bibnamefont{Nakai\textit{ et~al.}}},
\bibinfo{journal}{J. Phys. Soc. Jpn}
  \textbf{\bibinfo{volume}{77}}, \bibinfo{pages}{073701}
  (\bibinfo{year}{2008}).

\bibitem[{\citenamefont{Boeri {\textit{et~al.}}}(2008)\citenamefont{Boeri,
  Dolgov, and Golubov}}]{BoeriPRL2008}
\bibinfo{author}{\bibfnamefont{L.}~\bibnamefont{Boeri\textit{ et~al.}}},
\bibinfo{journal}{Phys. Rev. Lett.}
  \textbf{\bibinfo{volume}{101}}, \bibinfo{pages}{026403}
  (\bibinfo{year}{2008}).

\bibitem[{\citenamefont{Singh and Du}(2008)}]{SinghPRL2008}
\bibinfo{author}{\bibfnamefont{D.~J.~Singh and M.~-H.~Du}},
\bibinfo{journal}{Phys. Rev. Lett.}
  \textbf{\bibinfo{volume}{100}}, \bibinfo{pages}{237003}
  (\bibinfo{year}{2008}).

\bibitem[{\citenamefont{Mazin {\textit{et~al.}}}(2008)\citenamefont{Mazin,
  Singh, Johannes, and Du}}]{MazinPRL2008}
\bibinfo{author}{\bibfnamefont{I.~I.} \bibnamefont{Mazin\textit{ et~al.}}},
\bibinfo{journal}{Phys.~Rev.~Lett.}
  \textbf{\bibinfo{volume}{101}}, \bibinfo{pages}{057003}
  (\bibinfo{year}{2008}).

\bibitem[{\citenamefont{Kuroki {\textit{et~al.}}}(2008)\citenamefont{Kuroki,
  Onari, Arita, Usui, Tanaka, Kontani, and Aoki}}]{KurokiPRL2008}
\bibinfo{author}{\bibfnamefont{K.}~\bibnamefont{Kuroki\textit{ et~al.}}},
\bibinfo{journal}{Phys.~Rev.~Lett.}
  \textbf{\bibinfo{volume}{101}}, \bibinfo{pages}{087004}
  (\bibinfo{year}{2008}).

\bibitem[{Not({\natexlab{a}})}]{NoteCharacterization}
\bibinfo{note}{For more precise analysis of foreign phases in our samples, see
  Ref.~3 and its supplementary data.}

\bibitem[{\citenamefont{Cox and Sangster}(1985)}]{Vegard}
\bibinfo{author}{\bibfnamefont{A.~Cox and M.~J.~L.~Sangster}},
\bibinfo{journal}{J.~Phys.~C: Solid State
  Phys.} \textbf{\bibinfo{volume}{18}}, \bibinfo{pages}{L1123}
  (\bibinfo{year}{1985}).

\bibitem[{\citenamefont{Kim {\textit{et~al.}}}(2008)\citenamefont{Kim,
  Kamihara, Yoon, Han, Nomura, Matsuishi, Nakao, Tanabe, Hirano, and
  Hosono}}]{KimKamiharaJPSJ}
\bibinfo{author}{\bibfnamefont{S.~W.} \bibnamefont{Kim\textit{ et~al.}}},
\bibinfo{journal}{J.~Phys.~Soc.~Jpn.}
  \textbf{\bibinfo{volume}{77}}, \bibinfo{pages}{23} (\bibinfo{year}{2008}).

\bibitem[{\citenamefont{Mukuda {\textit{et~al.}}}(2008)\citenamefont{Mukuda,
  Terasaki, Kinouchi, Yashima, Kitaoka, Suzuki, Miyasaka, Tajima, Miyazawa,
  Shirage {\textit{et~al.}}}}]{MukudaJPSJ08}
\bibinfo{author}{\bibfnamefont{H.}~\bibnamefont{Mukuda\textit{ et~al.}}},
\bibinfo{journal}{J. Phys. Soc. Jpn.}
  \textbf{\bibinfo{volume}{77}}, \bibinfo{pages}{093704}
  (\bibinfo{year}{2008}).

\bibitem[{Not({\natexlab{b}})}]{Note}
\bibinfo{note}{At $x=0.07$, we observed a short component of $1/T_1$ below
  $\sim$40 K, but we could not ascribe the appearance of the short component to
  magnetic order because we observed neither broadening of $^{75}$As NMR
  spectra nor a resistive anomaly. A short component of $1/T_1$ is also
  observed below $\sim T_c$ for $x=0.11$. We speculate that the appearance of a
  short component of $1/T_1$ is due to residual disorder and/or vortex cores.}

\bibitem[{\citenamefont{Takeshita
  {\textit{et~al.}}}(2008)\citenamefont{Takeshita, Kadono, Hiraishi, Miyazaki,
  Koda, Kamihara, , and Hosono}}]{TakeshitaJPSJ08}
\bibinfo{author}{\bibfnamefont{S.}~\bibnamefont{Takeshita\textit{ et~al.}}},
\bibinfo{journal}{J.~Phys.~Soc.~Jpn.}
  \textbf{\bibinfo{volume}{77}}, \bibinfo{pages}{103703}
  (\bibinfo{year}{2008}).

\bibitem[{\citenamefont{Kohama {\textit{et~al.}}}(2008)\citenamefont{Kohama,
  Kamihara, Hirano, Kawaji, Atake, and Hosono}}]{KohamaFeAsPRB2008}
\bibinfo{author}{\bibfnamefont{Y.}~\bibnamefont{Kohama\textit{ et~al.}}},
\bibinfo{journal}{Phys. Rev. B}
  \textbf{\bibinfo{volume}{78}}, \bibinfo{pages}{020512(R)}
  (\bibinfo{year}{2008}).

\bibitem[{\citenamefont{Ding {\textit{et~al.}}}(2008)\citenamefont{Ding,
  Richard, Nakayama, Sugawara, Arakane, Sekiba, Takayama, Souma, Sato,
  Takahashi {\textit{et~al.}}}}]{HDingBaKFe2As2}
\bibinfo{author}{\bibfnamefont{H.}~\bibnamefont{Ding\textit{ et~al.}}},
\bibinfo{journal}{Europhys. Lett.}
  \textbf{\bibinfo{volume}{83}}, \bibinfo{pages}{47001} (\bibinfo{year}{2008}).

\bibitem[{\citenamefont{Hashimoto
  {\textit{et~al.}}}(2009)\citenamefont{Hashimoto, Shibauchi, Kato, Ikada,
  Okazaki, Shishido, Ishikado, Kito, Iyo, Eisaki
  {\textit{et~al.}}}}]{HashimotoPrFeAsO1-y}
\bibinfo{author}{\bibfnamefont{K.}~\bibnamefont{Hashimoto\textit{ et~al.}}},
\bibinfo{journal}{Phys.~Rev.~Lett.}
  \textbf{\bibinfo{volume}{102}}, \bibinfo{pages}{017002}
  (\bibinfo{year}{2009}).

\bibitem[{\citenamefont{Parker {\textit{et~al.}}}(2008)\citenamefont{Parker,
  Dolgov, Korshunov, Golubov, and Mazin}}]{ParkerPRB2008}
\bibinfo{author}{\bibfnamefont{D.}~\bibnamefont{Parker\textit{ et~al.}}},
\bibinfo{journal}{Phys.~Rev.~B}
  \textbf{\bibinfo{volume}{78}}, \bibinfo{pages}{134524}
  (\bibinfo{year}{2008}).

\bibitem[{\citenamefont{Chubukov
  {\textit{et~al.}}}(2008)\citenamefont{Chubukov, Efremov, and
  Eremin}}]{ChubukovPRB2008}
\bibinfo{author}{\bibfnamefont{A.~V.} \bibnamefont{Chubukov\textit{ et~al.}}},
\bibinfo{journal}{Phys.~Rev.~B}
  \textbf{\bibinfo{volume}{78}}, \bibinfo{pages}{134512}
  (\bibinfo{year}{2008}).

\bibitem[{\citenamefont{Parish {\textit{et~al.}}}(2008)\citenamefont{Parish,
  Hu, and Bernevig}}]{ParishPRB2008}
\bibinfo{author}{\bibfnamefont{M.~M.} \bibnamefont{Parish\textit{ et~al.}}},
\bibinfo{journal}{Phys.~Rev.~B}
  \textbf{\bibinfo{volume}{78}}, \bibinfo{pages}{144514}
  (\bibinfo{year}{2008}).

\bibitem[{\citenamefont{Bang and Choi}(2008)}]{Bang}
\bibinfo{author}{\bibfnamefont{Y.}~\bibnamefont{Bang\textit{ et~al.}}}
\bibinfo{journal}{Phys.~Rev.~B}
  \textbf{\bibinfo{volume}{78}}, \bibinfo{pages}{134523}
  (\bibinfo{year}{2008}).

\bibitem[{\citenamefont{Nagai {\textit{et~al.}}}(2008)\citenamefont{Nagai,
  Hayashi, Nakai, Nakamura, Okumura, and Machida}}]{NagaiNJP2008}
\bibinfo{author}{\bibfnamefont{Y.}~\bibnamefont{Nagai\textit{ et~al.}}},
\bibinfo{journal}{New~ J.~Phys.}
  \textbf{\bibinfo{volume}{10}}, \bibinfo{pages}{103026}
  (\bibinfo{year}{2008}).

\bibitem[{\citenamefont{Kitagawa
  {\textit{et~al.}}}(2008)\citenamefont{Kitagawa, Katayama, Ohgushi, Yoshida,
  and Takigawa}}]{KitagawaBaFe2As2}
\bibinfo{author}{\bibfnamefont{K.}~\bibnamefont{Kitagawa\textit{ et~al.}}},
\bibinfo{journal}{J.~Phys.~Soc.~Jpn.}
  \textbf{\bibinfo{volume}{77}}, \bibinfo{pages}{114709}
  (\bibinfo{year}{2008}).

\bibitem[{\citenamefont{Fukazawa
  {\textit{et~al.}}}(2008)\citenamefont{Fukazawa, Hirayama, Kondo, Yamazaki,
  Kohori, Takeshita, Miyazawa, Kito, Eisaki, and Iyo}}]{FukazawaJPSJ08}
\bibinfo{author}{\bibfnamefont{H.}~\bibnamefont{Fukazawa\textit{ et~al.}}},
\bibinfo{journal}{J.~Phys.~Soc.~Jpn.}
  \textbf{\bibinfo{volume}{77}}, \bibinfo{pages}{093706}
  (\bibinfo{year}{2008}).

\bibitem[{\citenamefont{Baek {\textit{et~al.}}}(2008)\citenamefont{Baek,
  Klimczuk, Ronning, Bauer, Thompson, and Curro}}]{BaekBaFe2As2}
\bibinfo{author}{\bibfnamefont{S.-H.} \bibnamefont{Baek\textit{ et~al.}}},
\bibinfo{journal}{Phys.~Rev.~B}
  \textbf{\bibinfo{volume}{78}}, \bibinfo{pages}{212509}
  (\bibinfo{year}{2008}).

\bibitem[{\citenamefont{Klingeler
  {\textit{et~al.}}}(2008)\citenamefont{Klingeler, Leps, Hellmann, Popa, Hess,
  Kondra, Hamann-Borrero, Behr, Kataev, and Buechner}}]{KlingelerLaFeAsOF}
\bibinfo{author}{\bibfnamefont{R.}~\bibnamefont{Klingeler\textit{ et~al.}}},
\bibinfo{journal}{arXiv:0808.0708}
  (\bibinfo{year}{2008}).

\bibitem[{\citenamefont{Takigawa
  {\textit{et~al.}}}(1991)\citenamefont{Takigawa, Reyes, Hammel, Thompson,
  Heffner, Fisk, and Ott}}]{TakigawaPRB1991}
\bibinfo{author}{\bibfnamefont{M.}~\bibnamefont{Takigawa\textit{ et~al.}}},
\bibinfo{journal}{Phys. Rev. B}
  \textbf{\bibinfo{volume}{43}}, \bibinfo{pages}{247} (\bibinfo{year}{1991}).

\bibitem[{\citenamefont{Terasaki
  {\textit{et~al.}}}(2009)\citenamefont{Terasaki, Mukuda, Yashima, Kitaoka,
  Miyazawa, Shirage, Kito, Eisaki, and Iyo}}]{TerasakiFeNMR}
\bibinfo{author}{\bibfnamefont{N.}~\bibnamefont{Terasaki\textit{ et~al.}}},
\bibinfo{journal}{J.~Phys.~Soc.~Jpn.}
  \textbf{\bibinfo{volume}{78}}, \bibinfo{pages}{013701}
  (\bibinfo{year}{2009}).

\bibitem[{\citenamefont{Ikeda}(2008)}]{IkedaJPSJ2008}
\bibinfo{author}{\bibfnamefont{H.}~\bibnamefont{Ikeda}},
  \bibinfo{journal}{J.~Phys.~Soc.~Jpn.} \textbf{\bibinfo{volume}{77}},
  \bibinfo{pages}{123707} (\bibinfo{year}{2008}).

\bibitem[{\citenamefont{Ohsugi {\textit{et~al.}}}(1994)\citenamefont{Ohsugi,
  Kitaoka, Ishida, and Asayama}}]{OhsugiJPSJ94}
\bibinfo{author}{\bibfnamefont{S.}~\bibnamefont{Ohsugi\textit{ et~al.}}},
\bibinfo{journal}{J. Phys. Soc. Jpn.}
  \textbf{\bibinfo{volume}{63}}, \bibinfo{pages}{700} (\bibinfo{year}{1994}).

\bibitem[{\citenamefont{Imai {\textit{et~al.}}}(2008)\citenamefont{Imai,
  Ahilan, Ning, McGuire, Sefat, Jin, Sales, and Mandrus}}]{ImaiProceedings}
\bibinfo{author}{\bibfnamefont{T.}~\bibnamefont{Imai\textit{ et~al.}}},
\bibinfo{journal}{J.~Phys.~Soc.~Jpn.}
  \textbf{\bibinfo{volume}{77}}, \bibinfo{pages}{Suppl.~C 47}
  (\bibinfo{year}{2008}).

\end{thebibliography}
\end{document}